\pdfoutput=1
\documentclass{article}
\usepackage[english]{babel}
\usepackage[utf8]{inputenc}
\usepackage{fontenc}
\usepackage{subcaption}

\usepackage{graphicx}
\usepackage{newclude}
\usepackage{amsfonts}
\usepackage{amsmath}
\usepackage{amssymb}
\usepackage{bbm}
\usepackage{float}
\usepackage{booktabs}
\usepackage{tablefootnote}
\usepackage{xcolor}
\usepackage{arydshln}
\usepackage{hyperref}

\usepackage{comment}

\graphicspath{{fig/}}

\begin{document}

\title{%
Experimental Study of Concise Representations of Concepts and Dependencies
}



\newcommand{\inst}[1]{$^{#1}$}
\newcommand{\orcidID}[1]{~\textsuperscript{\textit{#1}}}
\newtheorem{definition}{Def.}

\author{
Aleksey Buzmakov\inst{1}\orcidID{0000-0002-9317-8785} \\
Egor Dudyrev\inst{2}\orcidID{0000-0002-2144-3308} \\
Sergei O. Kuznetsov\inst{2}\orcidID{0000-0003-3284-9001}\\
Tatiana Makhalova\inst{3}\orcidID{0000-0002-6724-3803} \\
Amedeo Napoli\inst{3,1}\orcidID{0000-0001-5236-9561} \\
}
%
%
\date{
  HSE University, Perm, Russia \\
  HSE University, Moscow, Russia \\
  Université de Lorraine, CNRS, Inria, LORIA, F-54000 Nancy, France
}

\maketitle              


\begin{abstract}
In this paper we are interested in studying concise representations of concepts and dependencies, i.e., implications and association rules.
Such representations are based on equivalence classes and their elements, i.e., minimal generators, minimum generators including keys and passkeys, proper premises, and pseudo-intents.
All these sets of attributes are significant and well studied from the computational point of view, while their statistical properties remain to be studied.
This is the purpose of this paper to study these singular attribute sets and in parallel to study how to evaluate the complexity of a dataset from an FCA point of view.
In the paper we analyze the empirical distributions and the sizes of these particular attribute sets.
In addition we propose several measures of data complexity relying on these attribute sets in considering real-world and related randomized datasets.
\end{abstract}


%
\section{Introduction}
%


%
%

In this paper we are interested in measuring  ``complexity'' of a dataset  in terms of Formal Concept Analysis (FCA \cite{GanterW99}).
On the one hand, we follow the lines of \cite{MakhalovaBKN22} where the ``closure structure'' and the ``closure index'' are introduced and based on the so-called passkeys, i.e., minimum generators in an equivalence class of itemsets. On the other hand, we'd like to capture statistical properties of a dataset, not just extremal characteristics such as the size of a passkey.
The closure structure represents a dataset, so that closed itemsets are assigned to the level of the structure given by the size of their passkeys.
The complexity of the dataset can be read along the number of levels of the dataset and the distribution of itemsets w.r.t. frequency at each level.
The most interesting 
are the ``lower'' levels, i.e., the levels with the lowest closure index, as they usually contain itemsets with high frequency, contrasting the higher levels which contain itemsets with a quite low frequency.
Indeed, short minimum keys or passkeys correspond to implications in the related equivalence class with minimal left-hand side (LHS) and maximal right-hand side (RHS), which are the most informative implications \cite{PasquierBTL99,BastideTPSL00}.

Here we accept an alternative approach and we try to measure the complexity of a dataset in terms of five main elements that can be computed in a concept lattice, namely intents (closed sets), pseudo-intents, proper premises, keys (minimal generators), and passkeys (minimum generators).
We follow a more practical point of view and we study the distribution of these different elements in various datasets.
We also investigate the relations that these five elements have with one another, and the relations with implications and association rules.
For example, the number of intents gives the size of the lattice, while the number of pseudo-intents gives the size of the Duquenne-Guigues basis \cite{GuiguesD86}, and thus the size of the minimal implication basis representing the whole lattice. 
Moreover, passkeys are indicators related to the closure structure and the closure index indicates the number of levels in the structure. The size of the covering relation of the concept lattice gives the size of the ``base'' of association rules.  

Here we discuss alternative ways of defining the ``complexity'' of a dataset and how it can be measured in the related concept lattice that can be computed from this dataset.
For doing so, we introduce two main indicators, namely
(i) the probability that two concepts $C_1$ and $C_2$ are comparable,
(ii) given two intents $A$ and $B$, the probability that the union of these two intent is again an intent.
Both indicators are related to the distributivity of a lattice \cite{DaveyP90,Gratzer02}.
Indeed, a distributive lattice may appear as less complex than random lattices, since, given two intents $A$ and $B$, their meet $A \wedge B$ and their join $A \vee B$ are also intents.
Moreover, in a distributive lattice, all pseudo-intents are of size $1$,
meaning that Duquenne-Guigues implication base is very simple having premises of size 1.

Following the same idea, given a set of $n$ attributes, the Boolean lattice $\wp(n)$ is the largest lattice that one can build from a context of size $n\times n$, but $\wp(n)$ can also be considered as a simple lattice, since it can be represented by the set of its $n$ atoms, and moreover, the Dququenne-Guigues implication base is empty, so there are no nontrivial implications in this lattice.
In addition, a Boolean lattice is also distributive, thus it is simple in terms of the join of intents.

This is an original and practical study about the complexity of a dataset through an analysis of specific elements in the related concept lattice, namely intents, pseudo-intents, proper premises, keys, and passkeys.
Direct links are drawn with implications and association rules, making also a bridge between the present study in the framework of FCA, and approaches more related to data mining, actually pattern mining and association rule discovery.
Indeed, the covering relation of the concept lattice makes a concise representation of the set of association rules of the context~\cite{PasquierBTL99,BastideTPSL00}, so that every element of the covering relation, i.e., a pair of neighboring concepts or edge of the concept lattice, stays for an association rule, and reciprocally, every association rule can be given by a set of such edges.
Frequency distribution of confidence of the edges can be considered as an important feature of the lattice as a collection of association rules.

For studying practically this complexity, we have conducted a series of experiments where we measure the distribution of the different elements for real-world datasets and then for randomized datasets.
Actually randomized datasets are based on real-world datasets where either the distribution of crosses in columns is randomized or the whole set of crosses is randomized while keeping the density of the dataset.
We can observe that randomized datasets are usually more complex than real-world datasets. 
This means that, in general, the set of ``interesting elements'' in the lattice is smaller 
in real-world datasets.

The paper is organized as follows.
In the second section we introduce the theoretical background and necessary definitions.
Then the next section presents a range of experiments involving real-world and randomized datasets.
Finally, the results of experiments are discussed, and then we make a conclusion.


%
\section{Theoretical Background}
%


%
\subsection{Classes of Characteristic Attribute Sets}

Here we recall basic FCA definitions related to concepts, dependencies, and their minimal representations. After that we illustrate the definitions with a toy example.
Let us consider a formal context $K = (G, M, I)$ and prime operators:
\begin{align}
    A' &= \{m \in M \mid \forall g \in A: gIm \}, \quad A \subseteq G \\
    B' &= \{g \in G \mid \forall m \in B: gIm \}, \quad B \subseteq M
\end{align}

In what follows we illustrate the definitions using the ``four geometrical figures and their properties'' formal context presented in Table~\ref{tbl:toy_context} and introduced in \cite{KuznetsovO02}, where the set of objects $G = \{g_1 , g_2, g_3, g_4\}$ corresponds to \{equilateral triangle, rectangle triangle, rectangle, square\}), and the set of attributes $M = \{a, b, c, d\}$ corresponds to \{has 3 vertices, has 4 vertices, has a direct angle, equilateral\}.

\begin{table}
\centering
\begin{tabular}{r|ccccc}
\hline
{} &     a &      b &      c &      d  & e\\
\hline
$g_1$ & x &   &   & x & \\
$g_2$ & x &   & x &   & \\
$g_3$ &   & x & x &   & \\
$g_4$ &   & x & x & x & \\
\hline
\end{tabular}   
\medskip
\caption{The formal context of geometrical figures.}
\label{tbl:toy_context}
\end{table}

\begin{definition}[Intent or closed description]
A subset of attributes $B \subseteq M$ is an intent (is closed) iff $B'' = B$.
\end{definition}

In the running example (Table~\ref{tbl:toy_context}), $B = B'' = \{b, c\}$ is an intent and is the maximal subset of attributes describing the subset of objects $B' = \{g_2, g_3\}$.

\begin{definition}[Pseudo-intent]
A subset of attributes $P \subseteq M$  is a pseudo-intent iff:
\begin{enumerate}
    \item $P \neq P''$
    \item $Q'' \subset P$ for every pseudo-intent $Q \subset P$
\end{enumerate}
\end{definition}

Pseudo-intents are premises of implications of the cardinality-minimal implication basis called ``Duquenne-Guigues basis'' \cite{GuiguesD86} (DG-basis, also known as ``canonical basis'' or ``stembase'' \cite{GanterW99}).
In the current example (Table~\ref{tbl:toy_context}), the set of pseudo-intents is
$\big\{\{b\}, \{e\}, \{c, d\}, \{a,b,c\} \big\}$ since:
\begin{itemize}
    \item $\{b\}, \{e\}, \{c, d\}$ are minimal non-closed subsets of attributes, and
    \item $\{a,b,c\}$ is both non-closed and contains the closure $\{b, c\}$ of the pseudo-intent $\{b\}$.
\end{itemize} 

\begin{definition}[Proper premise]
A set of attributes $A \subseteq M$ is a proper premise iff:
$$A \cup \bigcup_{n \in A} \left( A \setminus \{n\} \right)'' \neq A''$$
\end{definition}

Proper premises are premises of so-called direct or proper-premise base of implications, from which one obtains all implications with a single application of Armstrong rules (see also~\cite{RysselDB14}).
In the running example (Table~\ref{tbl:toy_context}), $Q = \{a, b\}$ is a proper premise since the union of $Q$ with the closures of its subsets does not result in the closure of $Q$, i.e., $\{a, b\} \cup \{a\}'' \cup \{b\}'' = \{a, b\} \cup \{a\} \cup \{b,c\} = \{a,b,c\} \neq \{a,b,c,d,e\}$.

\begin{definition}[Generator]
A set of attributes $D \subseteq M$ is a generator iff $\exists B \subseteq M: D'' = B$.
\end{definition}

In this paper, every subset of attributes is a generator of a concept intent.
A generator is called non-trivial if it is not closed.
In the current example (Table~\ref{tbl:toy_context}), $D = \{a, b, d\}$ is a generator of $B = \{a,b,c,d,e\}$ since $B$ is an intent, $D \subseteq B$, and $D'' = B$.

\begin{definition}[Minimal generator, key]
A set of attributes $D \subseteq M$ is a minimal generator or a key of $D''$ iff
$ \nexists m \in D: (D \setminus \{m\})'' = D''$.
\end{definition}

A minimal generator is inclusion minimal in the equivalence class of subsets of attributes having the same closure \cite{PasquierBTL99,BastideTPSL00}.
Every proper premise is a minimal generator, however the converse does not hold in general.
In the current example (Table~\ref{tbl:toy_context}), $D = \{a, c, d\}$ is a minimal generator since none of its subsets $\{a, c\}, \{a, d\}, \{c, d\}$ generates the intent $D'' = \{a,b,c,d,e\}$.

\begin{definition}[Minimum generator, passkey]
A set of attributes $D \subseteq M$ is a minimum generator or a passkey iff $D$ is a minimal generator of $D''$ with the minimal size among all minimal generators of $D''$.
\end{definition}

A minimum generator (a passkey) is cardinality-minimal in the equivalence class of subsets of attributes having the same closure. 
In~\cite{MakhalovaBKN22} the size of a maximal passkey of a context was studied as an index of the context complexity. 
In the current example (Table~\ref{tbl:toy_context}), $D = \{b,d\}$ is a minimum generator of the intent $\{b,c,d\}$ since there is no other generator of smaller cardinality generating $D''$.
Meanwhile $D = \{a,c,d\}$ is not a minimum generator of $D'' = \{a,b,c,d,e\}$ since the subset $E = \{e\}$ has a smaller size and the same closure, i.e., $E''= D''$.

\begin{figure}[h!]
  \includegraphics[width=\textwidth]{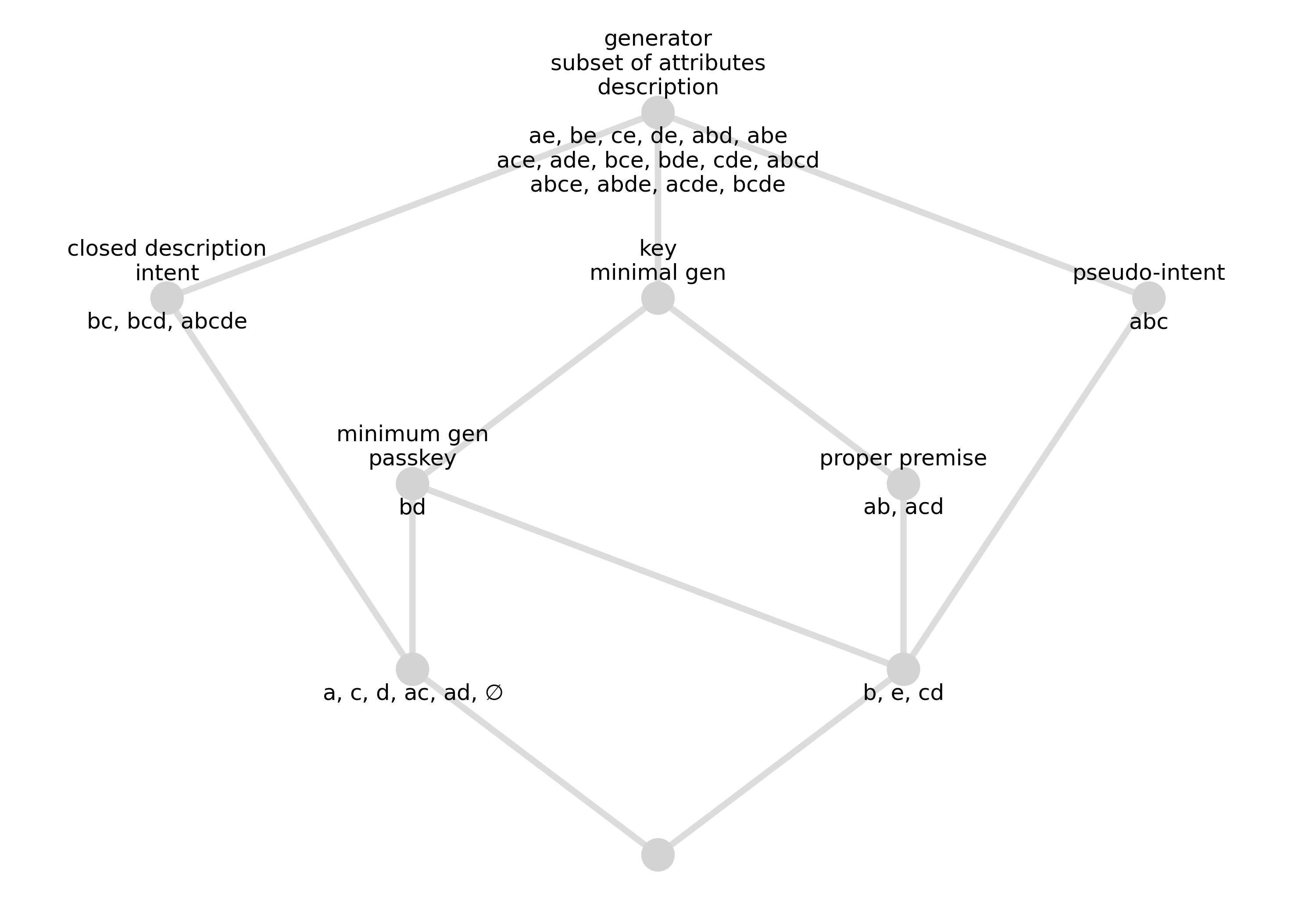}
  \caption{The concept lattice of ``characteristic attribute sets'' of the current context introduced in Table~\ref{tbl:toy_context}.
}
  \label{fig:descr_lattice_toy_context}
\end{figure}

Finally, we illustrate all these definitions at once.
To do so, we form the context of all possible classes of ``characteristic attribute sets'' of $M$ as they are introduced above, namely $(2^M, M_d, I_d)$, $M_d = \{\mathrm{passkey}, \mathrm{key}, \mathrm{proper\,premise},\ldots\}$, and $I_d$ defines whether a subset of attributes from $2^M$ belongs to a characteristic attribute set in $M_d$.
The concept lattice of this context is shown in Figure~\ref{fig:descr_lattice_toy_context}.

\subsection{Towards Measuring Data Complexity}

``Data complexity'' can mean many different things depending on particular problem of data analysis one wants to solve.
For example, data can be claimed to be complex when data processing takes a very long time, and this could be termed as ``computational complexity'' of data.
Alternatively, data can be considered as complex when data is hard to analyze and to interpret.
For example, it can be hard to apply FCA or machine learning algorithms, such as clustering, classification, or regression.
Accordingly, it is quite hard to define data complexity in general terms.


If we consider the dimension of interpretability, then the size of the elements to interpret and their number are definitely important elements.
In an ideal case, one would prefer a small number of ``elements'', say interesting subsets of attributes, to facilitate interpretation.
Indeed, less than five rules with a few attributes in the premises and in the conclusions are simpler to interpret than hundred of rules with more than ten attributes in the premises and conclusions.
Thus, it is natural to study how the number of elements are distributed w.r.t. their size.
Moreover, in most of the cases, large numbers of elements are associated with computational complexity.
Thus controlling the size and the number of elements is also a way to control computational complexity.

It should also be mentioned that the number of elements is related to the so-called ``VC-dimension''of a context~\cite{AlbanoC17}, which is the maximal size of a Boolean sublattice.
Accordingly, for our study about data complexity, we decided to count the number of concepts, pseudo-intents, proper premises, keys, and passkeys in order to understand and evaluate the complexity of data.
For all these elements, we also study the distribution of element sizes.

Additionally, we decided to measure the ``lattice complexity'' with two new measures which are related to what could be termed the ``linearity'' of the lattice.
Indeed, the most simple lattice structure that can be imagined is a chain, while the counterpart is represented by the Boolean lattice, i.e., the lattice with the largest amount of connections and concepts. 
However, it should be noticed that the Boolean lattice may be considered as complex from the point of view of interpretability, but very simple from the point of view of implication base, which is empty in such a lattice.

Then, a first way to measure the closeness of the lattice to a chain is the ``linearity index'' which is formally defined below as the probability that two random concepts are comparable in the lattice.

\begin{definition}
Given a lattice $\mathcal{L}$, the linearity index $\mathtt{LIN}(\mathcal{L})$ is defined as:
\begin{equation}
\mathtt{LIN}(\mathcal{L}) = \frac{1}{|\mathcal{L}|}\underset{c_1, c_2 \in \mathcal{L}, c_1 \neq c_2}{\sum} \mathbbm{1}(c_1 < c_2 \vee c_1 > c_2)
\end{equation}
where $|\mathcal{L}|$ denotes the number of the concepts in the lattice $\mathcal{L}$ and $\mathbbm{1}$ the indicator function which takes the value $1$ when the related constraint is true.
\end{definition}

The linearity index is maximal for a chain lattice and is minimal for the lattice of a nominal scale (or the lattice related to a bijection).
This index does not directly measure how well the lattice is interpretable.
One of the main interpretability properties is the size of some element sets, and in particular, the size and the structure of the implication basis.
One of the most simple structure for the implication basis can be found in distributive lattices, i.e. pseudo-intents are of size $1$.
Accordingly, we introduce the ``distributivity index'' which measures how a lattice is close to a distributive one.

\begin{definition}
Given a lattice $\mathcal{L}$, the distributivity index $\mathtt{DIST}(\mathcal{L})$ is defined as
\begin{equation}
\mathtt{DIST}(\mathcal{L}) = \frac{1}{|\mathcal{L}|}\underset{i_1,i_2 \in Intents(\mathcal{L}), i_1 \neq i_2}{\sum} \mathbbm{1}(i_1 \cup i_2 \in Intents(\mathcal{L})),
\end{equation}
where $Intents(\mathcal{L})$ is the set of concept intents in $\mathcal{L}$.
\end{definition}

The distributivity index is maximal for distributive lattices, and this includes chain lattices which are distributive lattices \cite{DaveyP90,Gratzer02}, and is again minimal for lattices of nominal scales which are not distributive.
However, it may sound strange to consider the lattices of nominal scales as complex.
Indeed they are simple from the viewpoint of implications.
For example, any pair of attributes from the lattice of a nominal scale --also termed as M3 for the most simple with 3 elements without top and bottom element-- can form the premise of an implication with a non-empty conclusion.
This indeed introduces many implications in the basis and this makes the DG-basis hard to interpret.

\subsection{Synthetic Complex Data}

In order to study different ways of measuring data complexity, we need to compare the behavior of different complexity indices for ``simple'' and ``complex'' data.
However, beforehand we cannot know which dataset is complex.
Accordingly, we will generate synthetic complex datasets and compare them with real-world data.
One way of generating complex data is ``randomization''.
Actually, randomized data cannot be interpreted since any possible result is an artifact of the method.
For randomized data we know beforehand that there cannot exist any rule or concept that have some meaning.
Thus, randomized data are good candidate data for being considered as ``complex''.

Now we discuss which randomization strategy should be used for generating such data.
A natural way is making randomized data similar in a certain sense to the real-world data they are compared to.
Firstly, when considering reference real-world data, it seems natural to keep the number of objects and attributes as they are in the real data.
Moreover, the ``density'' of the context, i.e., the number of crosses, significantly affects the size and the structure of the related concept lattice.
Thus, to ensure that randomized data are ``similar'' to the real data it is also natural to keep the density of data.
This gives us the first randomization strategy, i.e., for any real-world dataset we can generate a randomized dataset with the same number of objects and attributes, and with the same density.
Then, the crosses in the original context will be randomly spread along the new context in ensuring that the density is the same as in the original real data. 

For example, let us consider the context given in Table~\ref{tbl:descr_lattice_bobross}, where there are $8$ objects, $9$ attributes, and $35$ crosses.
Thus, any context with  $8$ objects, $9$ attributes, and $35$ crosses, can be considered as a randomization of this original context.
In our first randomization strategy, we suppose that the probability of generating any such randomized context is equally distributed.

\begin{table}
\begin{center}
\begin{tabular}{r|l|l|l|l|l|l|l|l|l}
\hline
descriptions & generator & closed & minimal & minimum & pseudo &  proper & key & passkey & intent \\
&        &  descr &     gen &     gen & intent & premise &  &      &     \\
\hline
67 &         x &      x &       x &       x &        &         &   x &       x &      x \\
45 &         x &      x &         &         &        &         &     &         &      x \\
41 &         x &        &       x &       x &      x &       x &   x &       x &        \\
125 &         x &        &       x &       x &        &       x &   x &       X &       \\
1 &         x &        &       x &         &      x &       x &   x &         &         \\
25 &         x &        &       x &         &        &       x &   x &         &        \\
33 &         x &        &         &         &      x &         &     &         &        \\
1048239 &         x &        &         &         &        &         &     &         &   \\
\hline
\end{tabular}
\medskip
\caption{%
The reduced context corresponding to the lattice of descriptions for Bob Ross dataset.}
\label{tbl:descr_lattice_bobross}
\end{center}
\end{table}

The randomized formal contexts for such strategy were studied in~\cite{BorchmannH16}.
The authors have found that the correlation between the number of concepts and the number of pseudo-intents has a non-random structure suggesting that fixing density is not enough in order to generate randomized data which are similar to the real one.
Accordingly, we also studied a randomization strategy that fixes the number of objects having a cross for every attribute as follows.
A randomized context is generated attribute by attribute.
The crosses in every column are randomly assigned while the number of crosses is not modified and remains the same as in the corresponding ``real attribute''.
This can be viewed as a permutation of the crosses within every column in the randomized context, a column being permuted independently of the others.
Such a procedure corresponds to the ``null hypothesis'' in statistical terms of independence between attributes.
Although such randomization strategy considers objects and attributes differently, it corresponds to typical cases in data analysis.
Indeed, in typical datasets, objects stand for observations that are described by different attributes.
The attributes correspond to any hypothesis of the data domain.
Then, analysts are usually interested in discovering some relations between attributes, and the hypothesis of attribute independence is a natural null hypothesis in such a setting.

For example, let as consider again the context in Table~\ref{tbl:descr_lattice_bobross}.
The numbers of objects and attributes in the randomized context remain the same.
Then a randomized context following the second strategy is any context having $8$ crosses for the first attribute, $2$ crosses for the second attribute, $5$ crosses for the third attribute, etc.
Finally, when randomizing data, one should have in mind that from a given real dataset many randomized datasets can be generated w.r.t. the same randomization strategy.
Thus, it is not enough to study only one random dataset for a given real dataset, but for the sake of objectivity, it is necessary to generate several randomized datasets and then to estimate the distribution of a characteristics under study within all randomized dataset.

In the next section we study different ways of measuring the complexity of a dataset and we observe that the complexity of randomized datasets is generally higher than the complexity of the corresponding real-world dataset.


%
\section{Experiments}
%


\subsection{Datasets}

For this preliminary study we selected $4$ small real-world datasets in order to support efficient computing of all necessary elements of the lattice.
Efficiency matters here because we involve randomization and computations are repeated hundreds of times for one dataset.
The study includes the following datasets:
``Live in water\footnote{\url{https://upriss.github.io/fca/examples.html}}'',
``Tea ladies\footnote{\url{https://upriss.github.io/fca/examples.html}}''
``Lattice of lattice properties\footnote{\url{https://upriss.github.io/fca/exampLes.html}}'', and
``Bob Ross episodes\footnote{\url{https://datahub.io/five-thirty-eight/bob-ross}}''.
For the sake of efficiency the fourth dataset was restricted to only first 20 attributes.

The datasets are analyzed in two different ways.
Firstly, the characteristic attribute sets are computed, e.g., concepts, keys, pseudo-intents, and then the relations existing between these elements are discussed. Secondly, we study the complexity of the datasets w.r.t. the number of characteristic attribute sets, and their distribution w.r.t. the size of these sets.
We also compared all these numerical indicators for real and the randomized datasets.

\subsection{Characteristic Attribute Sets in a Lattice}

In this section we study the relations between the different characteristic attribute sets.
The experiment pipeline reads as follows.
\begin{itemize}
\item
Given a context $K = (G,M,I)$, we compute all the possible ``attribute descriptions'', i.e., subsets of attributes in $2^M$, and we check whether a description shows some characteristic such as being closed, a pseudo-intent, a minimum generator, etc.
\item
We construct a new ``descriptions context'', namely $(2^M,CAS,I)$ where
$CAS = \{is\_closed, is\_gen, is\_key, is\_min\_gen, is\_passkey, is\_prop-prem\}$, 
and $I$ is the relation indicating that a given set of attributes has a characteristic in $Char$. 
\item
Finally, we construct the ``lattice of descriptions'' based on the description context''.
This lattice shows the relations existing between all generators -- subsets of attributes-- in a given dataset.
\end{itemize}

\begin{figure}[h!]
\includegraphics[width=\textwidth]{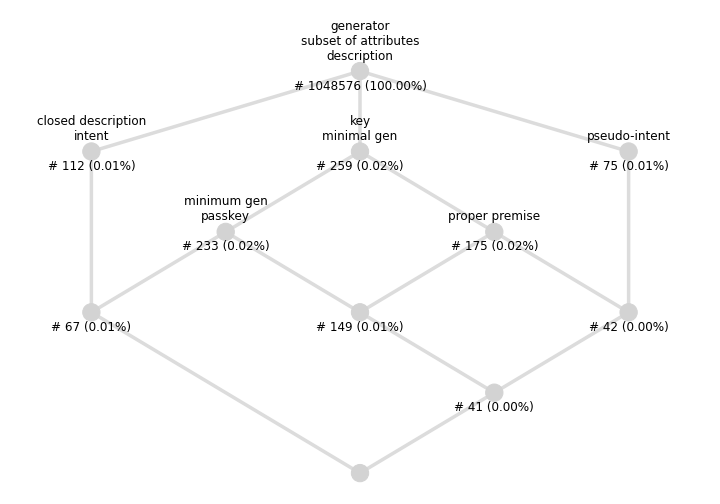}
\caption{The lattice of ``attribute descriptions'' for the ``Bob Ross'' dataset and
their distributions.}
 \label{fig:descr_lattice_bobross}
\end{figure}

The ``description context'' for the ``Bob Ross'' dataset is given in Table~\ref{tbl:descr_lattice_bobross} and the corresponding lattice is shown in Figure~\ref{fig:descr_lattice_bobross}.
From the lattice we can check that any two classes of descriptions may intersect if this is not forbidden by their definition (e.g., a description cannot be both an intent and a pseudo-intent).
Although such a lattice is computed for a particular dataset, this is the general lattice structure which is quite always obtained.
In some very small datasets it may happen that some characteristic attribute sets are missing.
For example, in the ``Live in water'' lattice the properties of being a key, being a passkey, and being a proper premise, all coincide and collapse into one single node. 


It is also very interesting to analyze the proportions of the sizes of classes of descriptions.
For example, in the ``Bob Ross'' context restricted to $20$ attributes, there are $2^{20}$ possible descriptions, but there are only $112$ of them which are closed, and only $259$ of them which are minimal generators.
Thus, the vast majority of the descriptions are ``useless'' in the sense that they do not correspond to any of the characteristic attribute subsets introduced above.
In the next subsection we consider the distributions of these characteristic attribute subsets.


%
\subsection{Data Complexity}

\begin{figure}
    \centering
    \begin{subfigure}{0.48\textwidth}
        \centering
        \includegraphics[width=\textwidth]{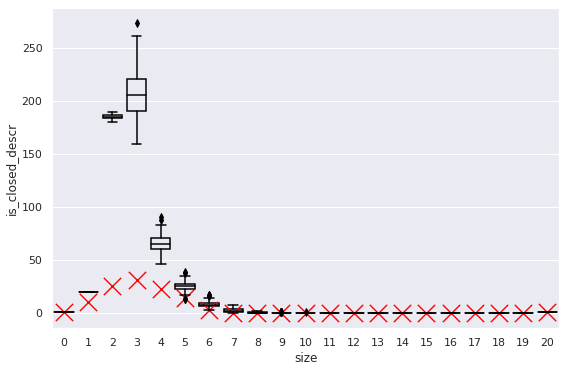}
        \caption{Intents}
        \label{fig:density-bob-intents}
    \end{subfigure}
    
    \vspace{1mm}
    
    \begin{subfigure}{0.48\textwidth}
        \centering
        \includegraphics[width=\textwidth]{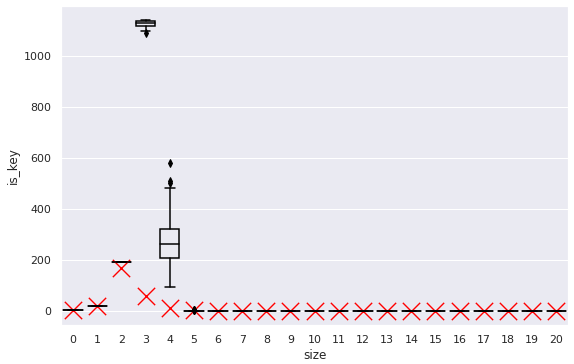}
        \caption{Keys}
        \label{fig:density-bob-key}
    \end{subfigure}
    \hfill
    \begin{subfigure}{0.48\textwidth}
        \centering
        \includegraphics[width=\textwidth]{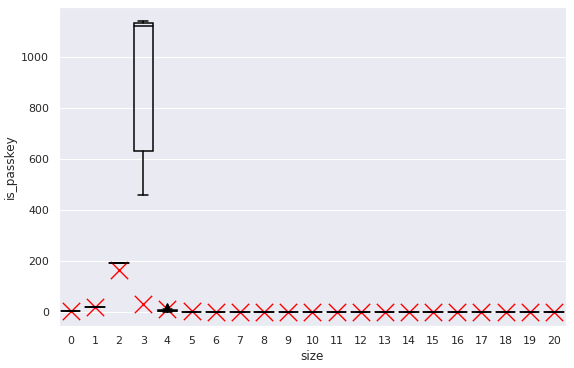}
        \caption{Passkeys}
        \label{fig:density-bob-passkey}
    \end{subfigure}
    
    \vspace{1mm}
    
    \begin{subfigure}{0.48\textwidth}
        \centering
        \includegraphics[width=\textwidth]{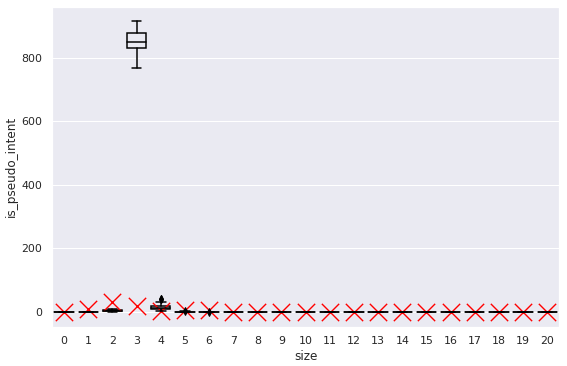}
        \caption{Pseudo-intents}
        \label{fig:density-bob-pseudo-i}
    \end{subfigure}
    \hfill
    \begin{subfigure}{0.48\textwidth}
        \centering
        \includegraphics[width=\textwidth]{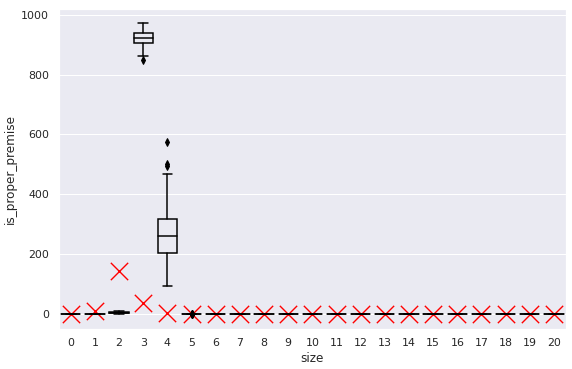}
        \caption{Proper premises}
        \label{fig:density-bob-proper-premises}
    \end{subfigure}
    
    \caption{The numbers of elements for the ``Bob Ross'' dataset w.r.t. context randomization based on density.}
\label{fig:density-bob}
\end{figure}

\begin{figure}
    \centering
    \begin{subfigure}{0.48\textwidth}
        \centering
        \includegraphics[width=\textwidth]{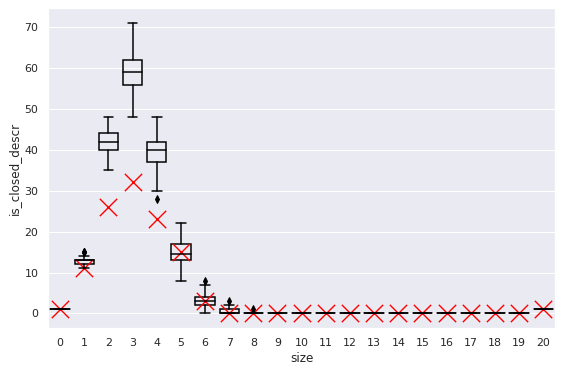}
        \caption{Intents}
        \label{fig:column-bob-intents}
    \end{subfigure}
    
    \vspace{1mm}
    
    \begin{subfigure}{0.48\textwidth}
        \centering
        \includegraphics[width=\textwidth]{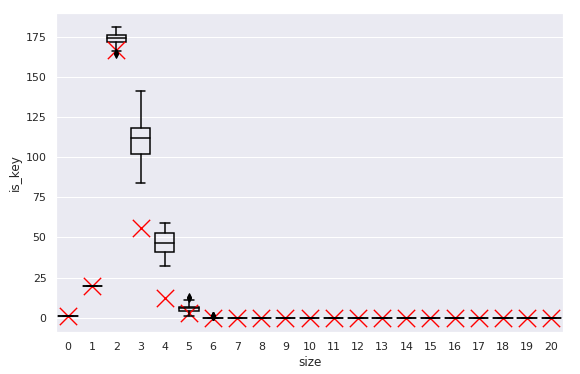}
        \caption{Keys}
        \label{fig:column-bob-key}
    \end{subfigure}
    \hfill
    \begin{subfigure}{0.48\textwidth}
        \centering
        \includegraphics[width=\textwidth]{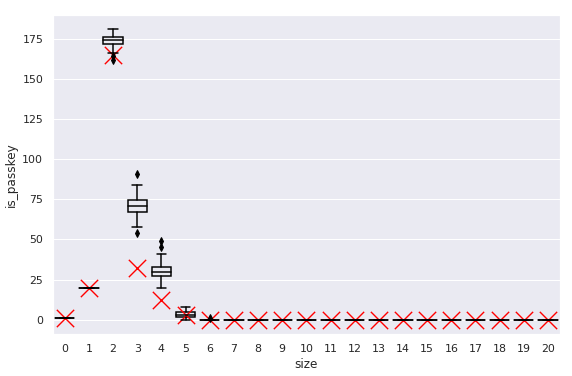}
        \caption{Passkeys}
        \label{fig:column-bob-passkey}
    \end{subfigure}
    
    \vspace{1mm}
    
    \begin{subfigure}{0.48\textwidth}
        \centering
        \includegraphics[width=\textwidth]{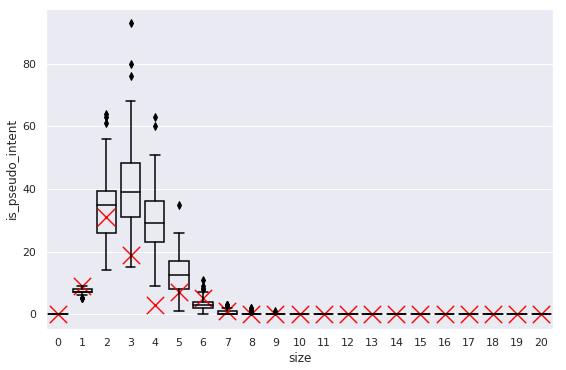}
        \caption{Pseudo-intents}
        \label{fig:column-bob-pseudo-i}
    \end{subfigure}
    \hfill
    \begin{subfigure}{0.48\textwidth}
        \centering
        \includegraphics[width=\textwidth]{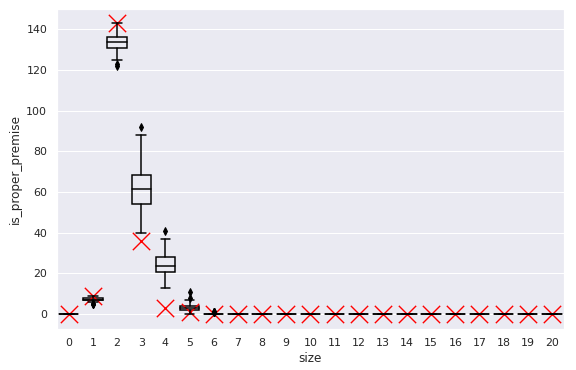}
        \caption{Proper premises}
        \label{fig:column-bob-proper-premises}
    \end{subfigure}
    
    \caption{The numbers of elements for the ``Bob Ross'' dataset w.r.t. context randomization based on column-wise permutations.}
    \label{fig:column-bob}
\end{figure}

For analyzing data complexity, we start by comparing the numbers of elements in real data and in randomized data.
In Figure~\ref{fig:density-bob-intents}~and~\ref{fig:column-bob} the distributions of different lattice elements for ``Bob Ross'' dataset is shown\footnote{All figures are in supplementary materials \url{https://yadi.sk/i/8_5EEvY4zNi82g}}.
Along the horizontal axis the sizes of the elements are shown, i.e., the number of attributes in the intent, pseudo-intent, key, etc.
Along the vertical axis the number of elements of the corresponding sizes are shown.

Red crosses shows the values corresponding for real data and the boxplots visualize the values found in random data.
There were $100$ randomizations and thus boxplots are based on these $100$ values.
A box corresponds to the 50\% middle values among $100$ values.
In addition, it should be noticed that these two figures differ in the randomization strategy.
Figure~\ref{fig:density-bob-intents} corresponds to density-based randomization while Figure~\ref{fig:column-bob} shows randomization based on column permutations.

\begin{figure}
    \centering
    \begin{subfigure}{0.48\textwidth}
        \centering
        \includegraphics[width=\textwidth]{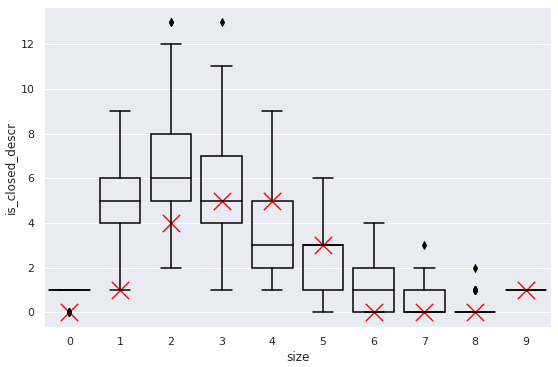}
        \caption{Intents}
        \label{fig:density-water-intents}
    \end{subfigure}
    
    \vspace{1mm}
    
    \begin{subfigure}{0.48\textwidth}
        \centering
        \includegraphics[width=\textwidth]{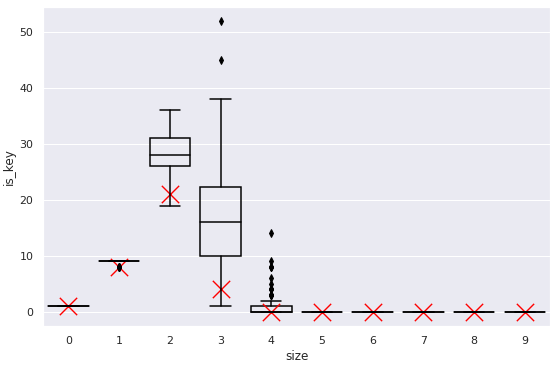}
        \caption{Keys}
        \label{fig:density-water-key}
    \end{subfigure}
    \hfill
    \begin{subfigure}{0.48\textwidth}
        \centering
        \includegraphics[width=\textwidth]{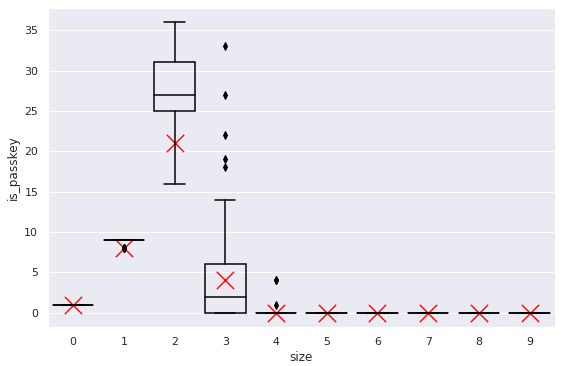}
        \caption{Passkeys}
        \label{fig:density-water-passkey}
    \end{subfigure}
    
    \vspace{1mm}
    
    \begin{subfigure}{0.48\textwidth}
        \centering
        \includegraphics[width=\textwidth]{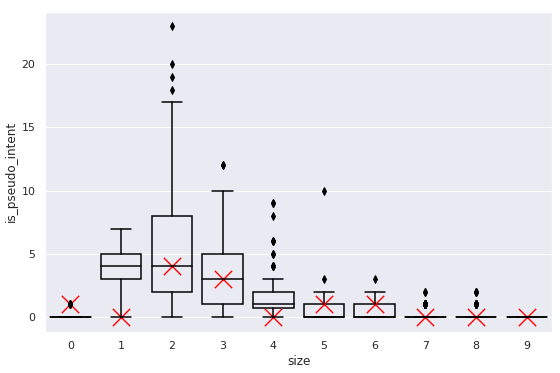}
        \caption{Pseudo-intents}
        \label{fig:density-water-pseudo-i}
    \end{subfigure}
    \hfill
    \begin{subfigure}{0.48\textwidth}
        \centering
        \includegraphics[width=\textwidth]{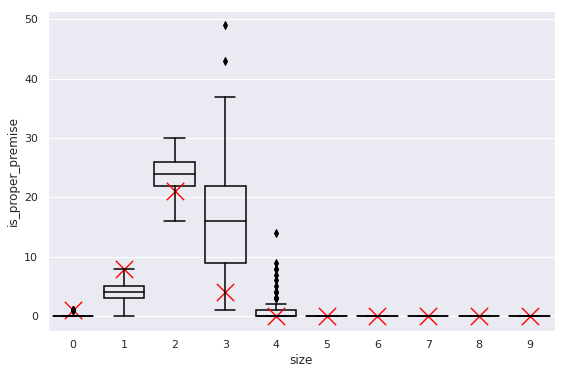}
        \caption{Proper premises}
        \label{fig:density-water-proper-premises}
    \end{subfigure}
    
    \caption{The numbers of elements for the ``Live in water'' dataset w.r.t. context randomization based on density.}
    \label{fig:density-water}
\end{figure}

From both figures we can observe that randomized data contain significantly larger numbers of elements than the real data.
Moreover, the sizes of the elements for randomized data are larger than the sizes for real data.
Similar figures can be built for ``Tea Ladies'' and ``Lattice of lattice properties'' datasets.
However, we cannot distinguish the real dataset and the randomized data for the ``Live in Water'' dataset  (see Figure~\ref{fig:density-water}).
This can be explained by the fact that either the dataset does not contain deep dependencies, or the dataset is too small, i.e., the randomized dataset cannot be substantially different from the original real one.
%

\begin{figure}
    \centering
    
    Density based randomization
    
    \begin{subfigure}{0.24\textwidth}
        \centering
        \includegraphics[width=\textwidth]{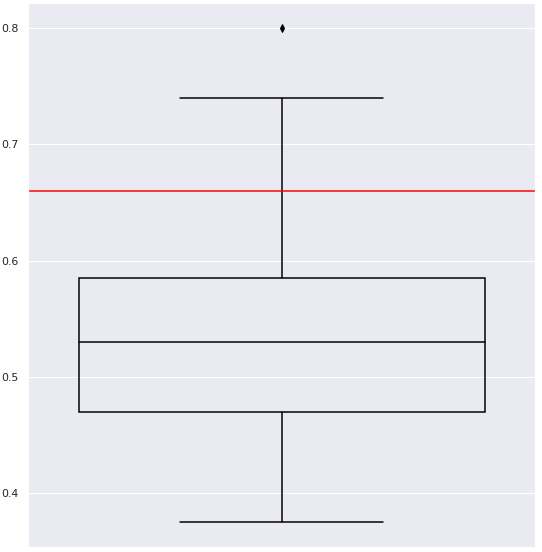}
        \caption{Live in water}
        \label{fig:density-water-lin}
    \end{subfigure}
    \hfill
    \begin{subfigure}{0.24\textwidth}
        \centering
        \includegraphics[width=\textwidth]{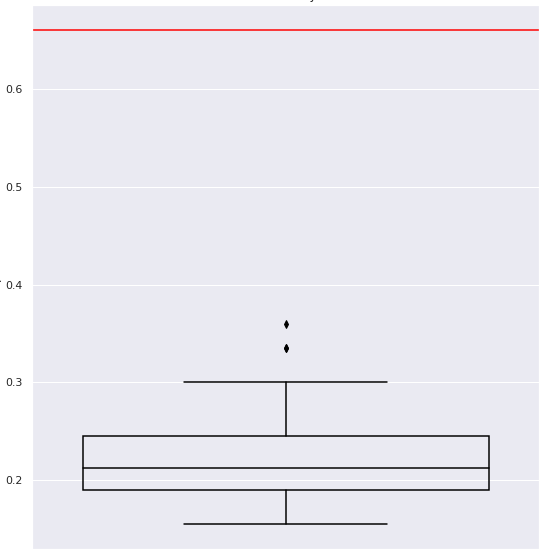}
        \caption{Lattice}
        \label{fig:density-lattice-lin}
    \end{subfigure}
    \hfill
    \begin{subfigure}{0.24\textwidth}
        \centering
        \includegraphics[width=\textwidth]{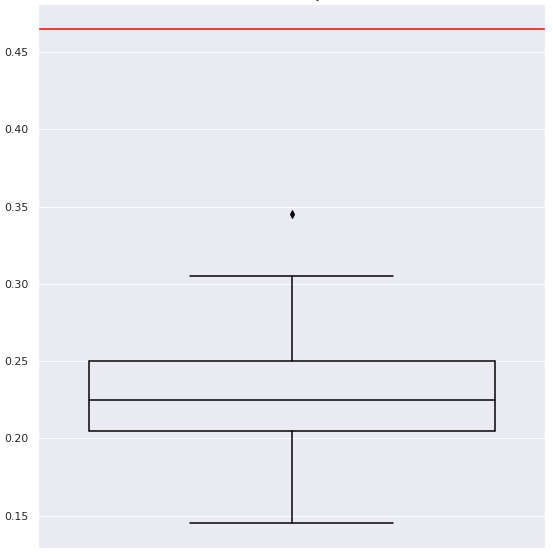}
        \caption{Tea Lady}
        \label{fig:density-tea-lin}
    \end{subfigure}
    \hfill
    \begin{subfigure}{0.24\textwidth}
        \centering
        \includegraphics[width=\textwidth]{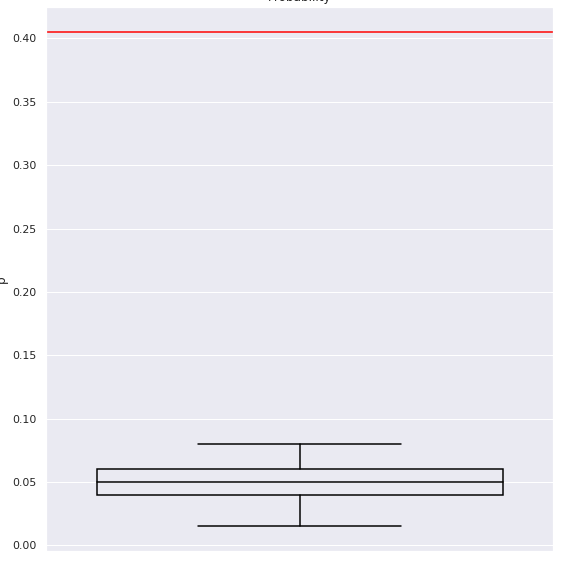}
        \caption{Bob Ross}
        \label{fig:density-bob-lin}
    \end{subfigure}
    
    \vspace{1mm}
    Randomization based on column permutations

    \begin{subfigure}{0.24\textwidth}
        \centering
        \includegraphics[width=\textwidth]{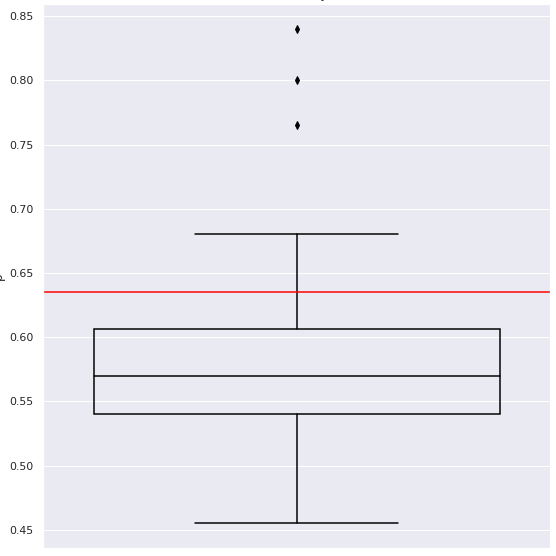}
        \caption{Live in water}
        \label{fig:column-water-lin}
    \end{subfigure}
    \hfill
    \begin{subfigure}{0.24\textwidth}
        \centering
        \includegraphics[width=\textwidth]{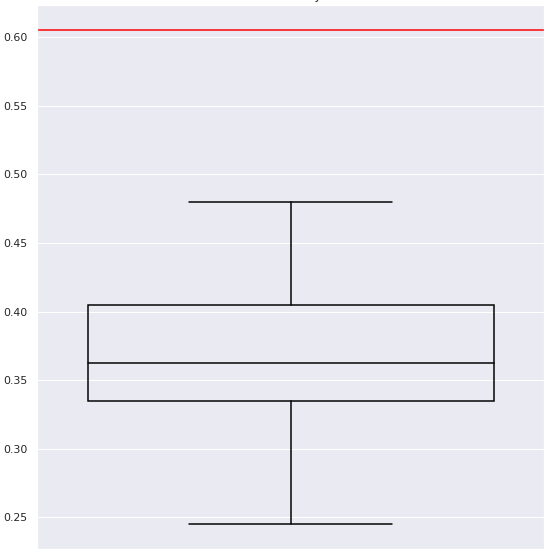}
        \caption{Lattice}
        \label{fig:column-lattice-lin}
    \end{subfigure}
    \hfill
    \begin{subfigure}{0.24\textwidth}
        \centering
        \includegraphics[width=\textwidth]{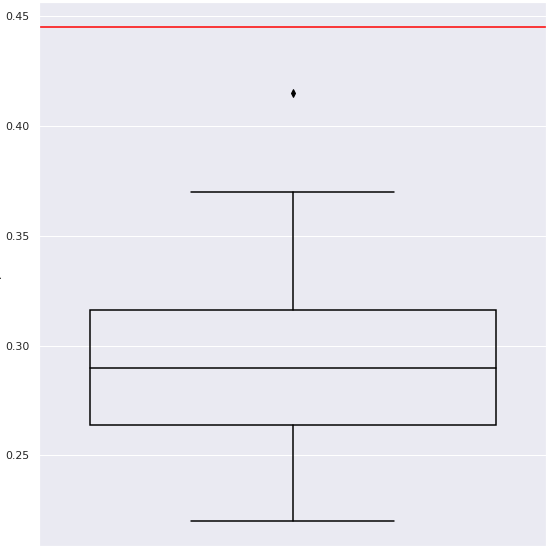}
        \caption{Tea Lady}
        \label{fig:column-tea-lin}
    \end{subfigure}
    \hfill
    \begin{subfigure}{0.24\textwidth}
        \centering
        \includegraphics[width=\textwidth]{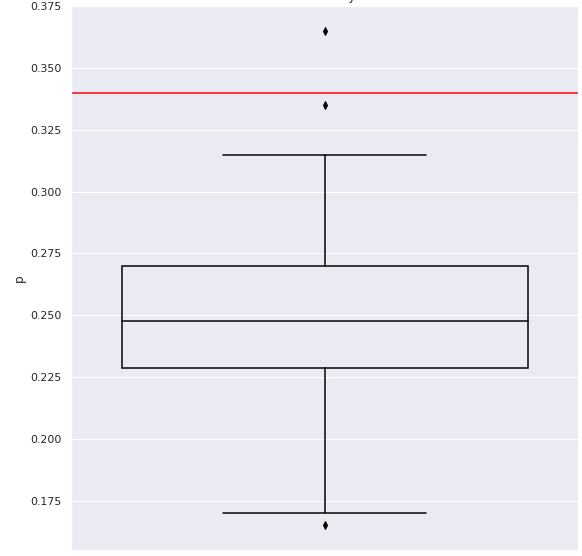}
        \caption{Bob Ross}
        \label{fig:column-bob-lin}
    \end{subfigure}
   
    \caption{The distributivity index for different datasets and different randomizations.}
    \label{fig:distributivity}
\end{figure}

\begin{figure}
    \centering
    
    Density based randomization
    
    \begin{subfigure}{0.24\textwidth}
        \centering
        \includegraphics[width=\textwidth]{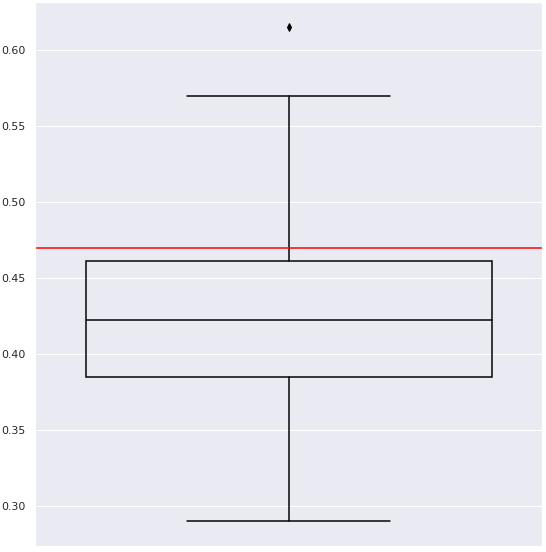}
        \caption{Live in water}
        \label{fig:density-water-lin}
    \end{subfigure}
    \hfill
    \begin{subfigure}{0.24\textwidth}
        \centering
        \includegraphics[width=\textwidth]{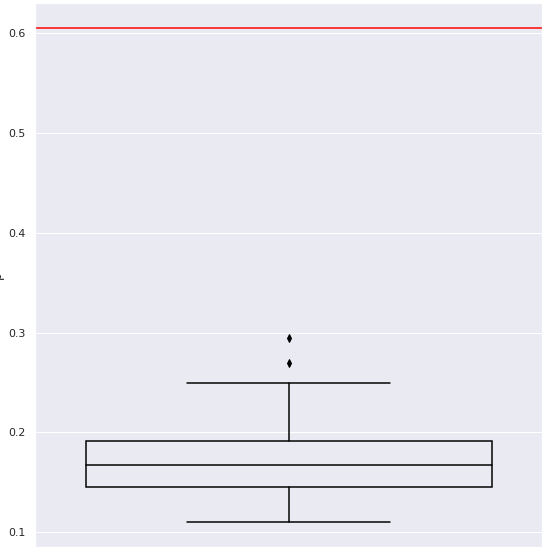}
        \caption{Lattice}
        \label{fig:density-lattice-lin}
    \end{subfigure}
    \hfill
    \begin{subfigure}{0.24\textwidth}
        \centering
        \includegraphics[width=\textwidth]{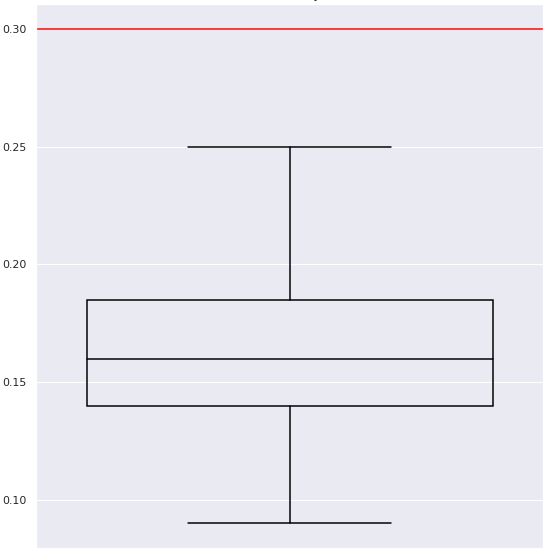}
        \caption{Tea Lady}
        \label{fig:density-tea-lin}
    \end{subfigure}
    \hfill
    \begin{subfigure}{0.24\textwidth}
        \centering
        \includegraphics[width=\textwidth]{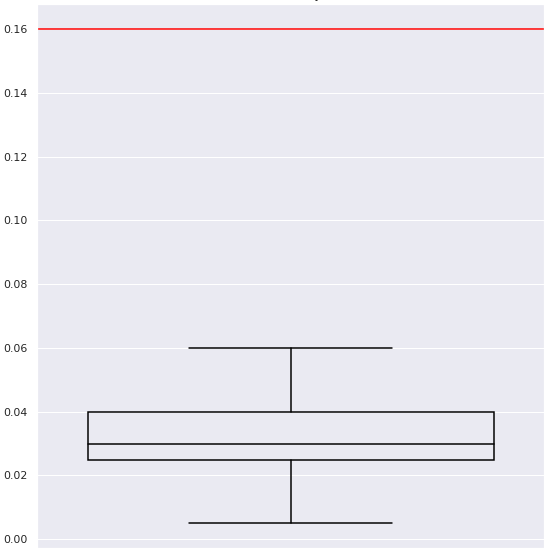}
        \caption{Bob Ross}
        \label{fig:density-bob-lin}
    \end{subfigure}
    
    \vspace{1mm}
    Randomization based on column permutations

    \begin{subfigure}{0.24\textwidth}
        \centering
        \includegraphics[width=\textwidth]{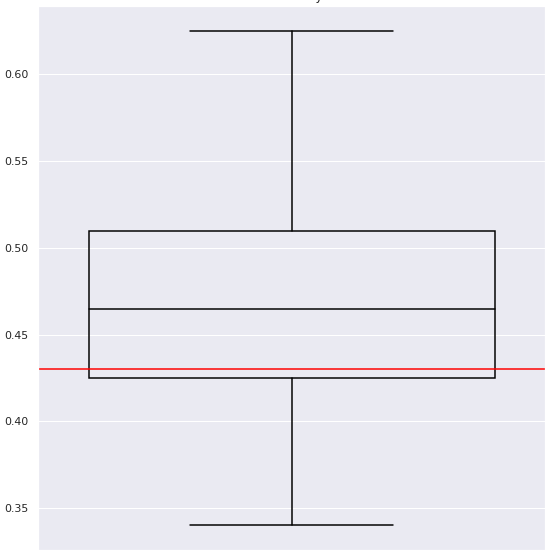}
        \caption{Live in water}
        \label{fig:column-water-lin}
    \end{subfigure}
    \hfill
    \begin{subfigure}{0.24\textwidth}
        \centering
        \includegraphics[width=\textwidth]{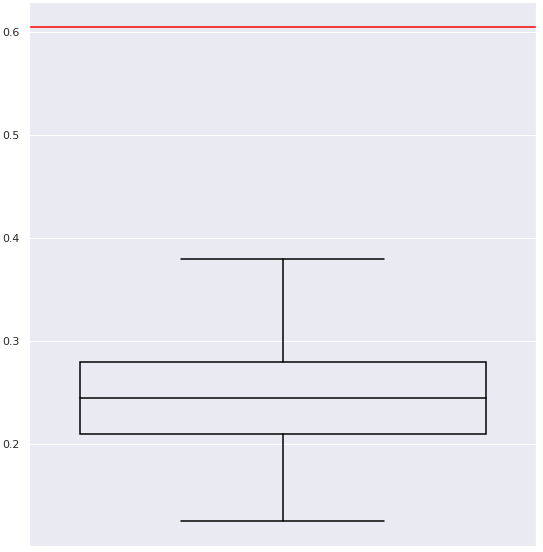}
        \caption{Lattice}
        \label{fig:column-lattice-lin}
    \end{subfigure}
    \hfill
    \begin{subfigure}{0.24\textwidth}
        \centering
        \includegraphics[width=\textwidth]{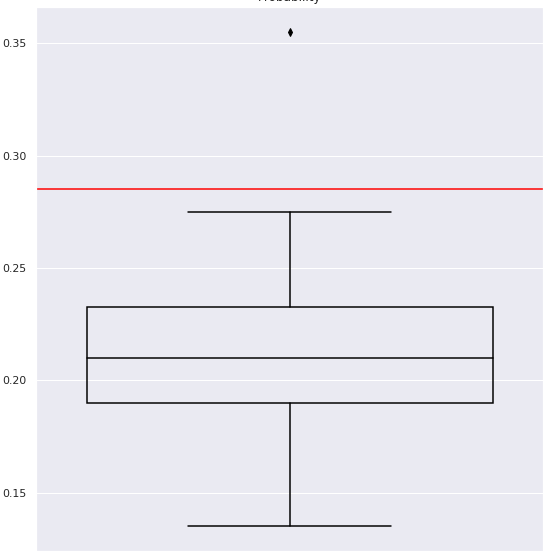}
        \caption{Tea Lady}
        \label{fig:column-tea-lin}
    \end{subfigure}
    \hfill
    \begin{subfigure}{0.24\textwidth}
        \centering
        \includegraphics[width=\textwidth]{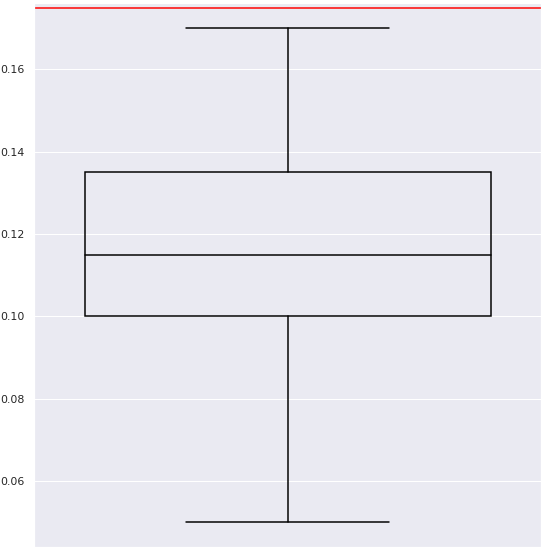}
        \caption{Bob Ross}
        \label{fig:column-bob-lin}
    \end{subfigure}
   
    \caption{The linearity index for different datasets and different randomizations.}
    \label{fig:linearity}
\end{figure}

Let us now study how the linearity index and the distributivity index measure the complexity of a dataset.
Figures~\ref{fig:linearity} and~\ref{fig:distributivity} show the values of the linearity and distributivity indices correspondingly w.r.t. different randomizations. 
From these figure we can see that datasets built from density-based randomization are more different from the real datasets than the randomized datasets built from column-wise permutations.
We should also notice that the values of the linearity and distributivity indices show a substantial dependence w.r.t. density of the corresponding context.
Indeed, if we look at the randomized datasets, we can see that the distributions of the linearity and distributivity indices are different.
This can be explained either by the context density and by the context size.
Thus, we cannot have any reference values for these indices that would split between ``complex'' and not ``simple'' data.
However, comparing the values of the index to the distribution of these indices allow one to decide on complexity of the data.
Finally, in all datasets but ``Live in water'' both indices, linearity and distributivity, have higher values for real datasets than for randomized datasets.
This shows again that real datasets are more structured than their randomized counterparts.



%
\section{Conclusion}

In this paper we have introduced and studied ``concise representations'' of datasets given by related contexts and concept lattices, and characteristic attributes sets based on equivalence classes, i.e., minimal generators, minimum generators including keys and passkeys, proper premises, and pseudo-intents.
We have also introduced two new indices for measuring the complexity of a datatet, the linearity index for checking the direct dependencies between concepts or how a concept lattice is close to a chain, and the distributivity lattice which measures how close is a concept lattice to a distributive lattice (where all pseudo-intents are of length $1$, thus leading to sets of simple implications).
We have also proposed a series of experiments where we analyze real-world datasets and their randomized counterparts.
As expected, the randomized datasets are more complex that the real ones.

The future work will be to improve this study in several directions, by studying more deeply the role of both indices, the linearity index and the distributivity index, by analyzing more larger datasets, and more importantly by analyzing the complexity from the point of view of the generated implications and association rules.
This is a first step in this direction and we believe that FCA can bring a substantial support for analyzing data complexity in general.

%
%

%
\bibliographystyle{plain}
\bibliography{bibliography}

\begin{thebibliography}{10}

\bibitem{AlbanoC17}
Alexandre Albano and Bogdan Chornomaz.
\newblock {Why concept lattices are large: extremal theory for generators,
  concepts, and VC-dimension}.
\newblock {\em {International Journal of General Systems}}, 46(5):440--457,
  2017.

\bibitem{BastideTPSL00}
Yves Bastide, Rafik Taouil, Nicolas Pasquier, Gerd Stumme, and Lotfi Lakhal.
\newblock Mining frequent patterns with counting inference.
\newblock {\em SIGKDD Exploration Newsletter}, 2(2):66--75, 2000.

\bibitem{BorchmannH16}
Daniel Borchmann and Tom Hanika.
\newblock {Some Experimental Results on Randomly Generating Formal Contexts}.
\newblock In Marianne Huchard and Sergei~O. Kuznetsov, editors, {\em
  {Proceedings of the 13th International Conference on Concept Lattices and
  Their Applications (CLA)}}, volume 1624 of {\em {CEUR} Workshop Proceedings},
  pages 57--69. CEUR-WS.org, 2016.

\bibitem{DaveyP90}
Brian~A. Davey and Hilary~A. Priestley.
\newblock {\em {Introduction to Lattices and Order}}.
\newblock Cambridge University Press, Cambridge, UK, 1990.

\bibitem{GanterW99}
Bernhard Ganter and Rudolf Wille.
\newblock {\em {Formal Concept Analysis}}.
\newblock Springer, Berlin, 1999.

\bibitem{Gratzer02}
George Gr{\"a}tzer.
\newblock {\em {General Lattice Theory (Second Edition)}}.
\newblock Birk{\"a}uzer, 2002.

\bibitem{GuiguesD86}
Jean-Luc Guigues and Vincent Duquenne.
\newblock Famile minimale d'implications informatives resultant d'un tableau de
  donn\'ees binaire.
\newblock {\em Mathematique, Informatique et Sciences Humaines}, 95:5--18,
  1986.

\bibitem{KuznetsovO02}
Sergei~O. Kuznetsov and Sergei~A. Obiedkov.
\newblock {Comparing performance of algorithms for generating concept
  lattices}.
\newblock {\em {Journal of Experimental \& Theoretical Artificial
  Intelligence}}, 14(2/3):189--216, 2002.

\bibitem{MakhalovaBKN22}
Tatiana Makhalova, Aleksey Buzmakov, Sergei~O. Kuznetsov, and Amedeo Napoli.
\newblock {Introducing the closure structure and the GDPM algorithm for mining
  and understanding a tabular datasets}.
\newblock {\em International Journal of Approximate Reasoning}, 145:75--90,
  2022.

\bibitem{PasquierBTL99}
Nicolas Pasquier, Yves Bastide, Rafik Taouil, and Lotfi Lakhal.
\newblock {Pruning Closed Itemset Lattices for Association Rules}.
\newblock {\em International Journal of Information Systems}, 24(1):25--46,
  1999.

\bibitem{RysselDB14}
Uwe Ryssel, Felix Distel, and Daniel Borchmann.
\newblock {Fast algorithms for implication bases and attribute exploration
  using proper premises}.
\newblock {\em {Annals of Mathematics and Artificial Intelligence}},
  70(1-2):25--53, 2014.

\end{thebibliography}
\newpage
\section{Appendix: Figures Related to Experiments}
%

\newcommand{\commment}[1]{}
\commment{
\begin{figure}[H]
    \centering
    \includegraphics[width=\textwidth]{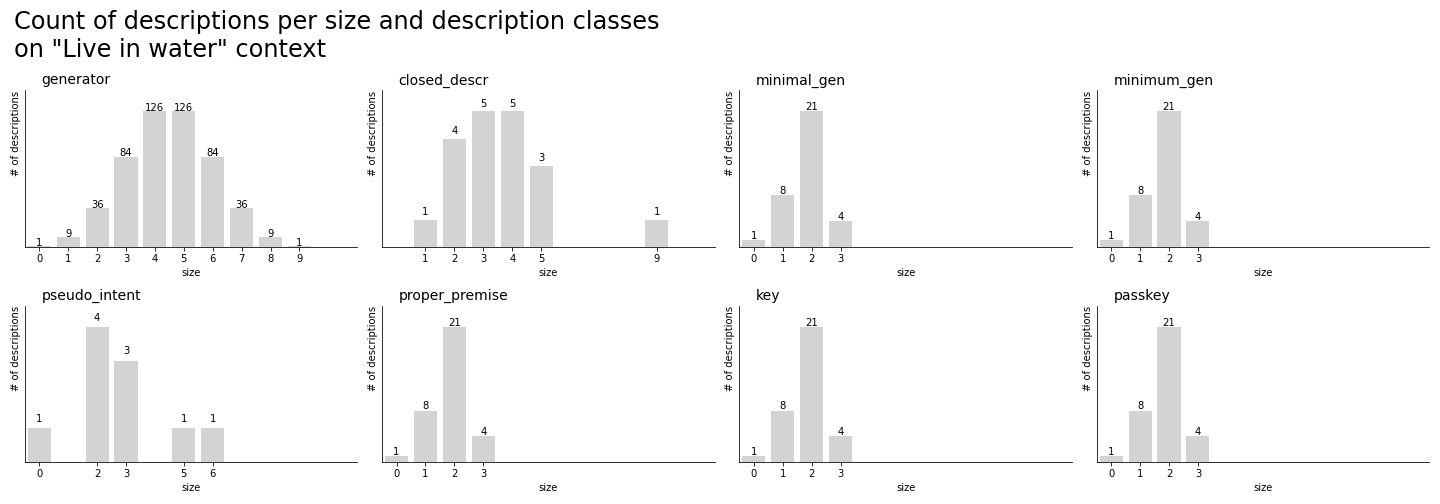}
    \caption{Histograms for "Live in water" dataset}
    \label{fig:hist_water}
\end{figure}

\begin{figure}[H]
    \centering
    \includegraphics[width=\textwidth]{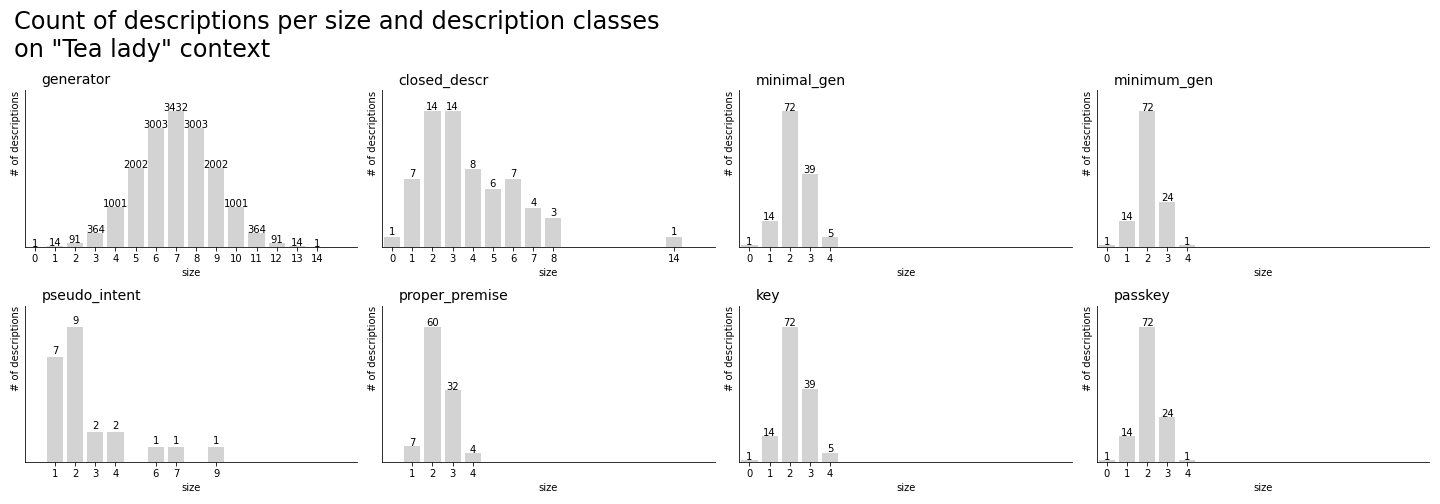}
    \caption{Histograms for "Tea ladies" dataset}
    \label{fig:hist_tealadies}
\end{figure}

\begin{figure}[H]
    \centering
    \includegraphics[width=\textwidth]{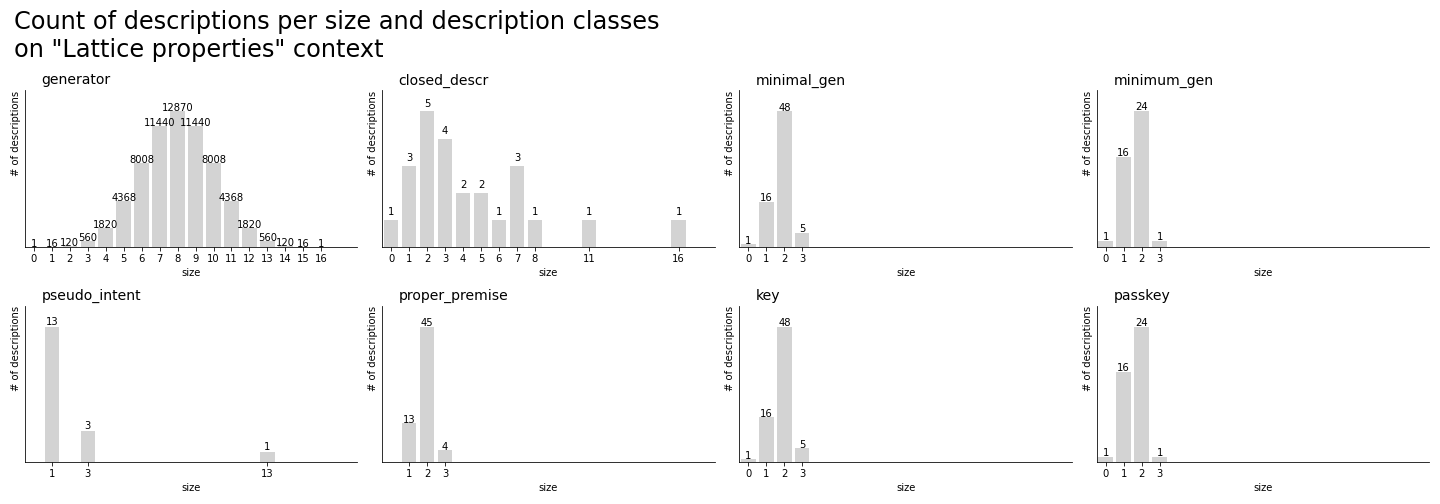}
    \caption{Histograms for "Lattice of lattice properties" dataset}
    \label{fig:hist_lattices}
\end{figure}

\begin{figure}[H]
    \centering
    \includegraphics[width=\textwidth]{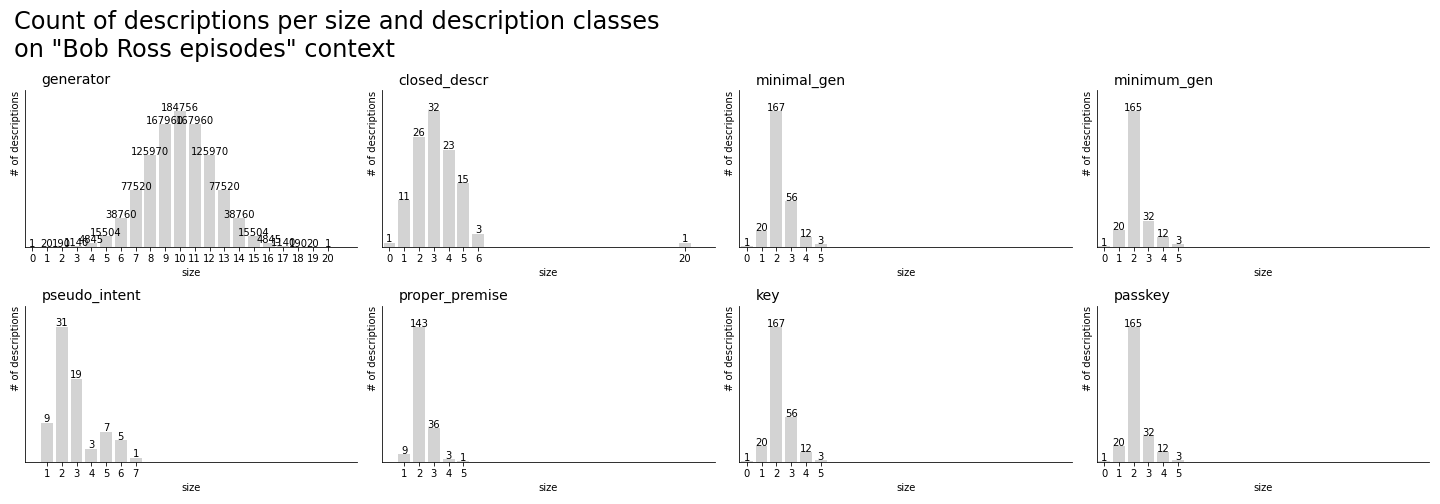}
    \caption{The histograms for ``Bob Ross'' dataset includng only the $20$ first attributes.}
    \label{fig:hist_bobross}
\end{figure}
}

\end{document}